\documentclass{article}
\usepackage{emulateapj}
\usepackage{times}

\usepackage{psfig}

\newcommand\rozanska{R\'o$\dot{\rm z}$a\'nska }
\newcommand\zycki{$\dot{\rm Z}$ycki }

\newcommand\fx{F_{\rm x}}

\newcommand\dm{\dot{m}}
\newcommand\fdisk{F_{\rm d}}

\newcommand\teff{T_{\rm eff}}

\def\>{$>$}
\def\<{$<$}

\def\simlt{\lower.5ex\hbox{$\; \buildrel < \over \sim \;$}}
\def\simgt{\lower.5ex\hbox{$\; \buildrel > \over \sim \;$}}
\def\sqr#1#2{{\vcenter{\hrule height.#2pt
      \hbox{\vrule width.#2pt height#1pt \kern#1pt
         \vrule width.#2pt}
      \hrule height.#2pt}}}

\begin{document}

\title{Ionization physics evidence for magnetic flare origin of the
X-rays in AGN}

\author{Sergei Nayakshin\altaffilmark{1}}

\affil{NASA/GSFC, LHEA, Code 661, Greenbelt, MD, 20771}
\altaffiltext{1}{National Research Council Associate}

\begin{abstract}
We present full disk X-ray reflection spectra for two currently
popular accretion flow geometries for AGN -- the lamppost model
frequently used to discuss the iron line reverberation in AGN, and the
model where the X-rays are produced in magnetic flares above a cold
accretion disk (AD). The lamppost spectra contain several
spectroscopic features characteristic of highly ionized material that
are not seen in the X-ray spectra of most AGN. The magnetic flare
model, on the other hand, produces reflected spectra that are roughly
a super-position of a power-law and a {\em neutral-like} reflection
and iron K$\alpha$ line, and are thus more in line with typical AGN
X-ray spectra. Furthermore, because of the difference in the
ionization structure of the illuminated material in the two models,
the line equivalent width increases with the X-ray luminosity, $L_x$,
for the lamppost, and decreases with $L_x$ for the flare model. In
light of these theoretical insights, recent iron line reverberation
studies of AGN, the X-ray Baldwin effect, and the general lack of
X-ray reflection features in distant quasars all suggest that, for
high accretion rates, the cold accretion disk is covered by a Thomson
thick, {\em completely ionized} skin. Because the latter is only
possible when the X-rays are concentrated to small emitting regions,
we believe that this presents a strong evidence for the magnetic flare
origin of X-rays in AGN.
\end{abstract}


\section{Introduction}\label{sect:intro}

Iron K$\alpha$ emission and the so-called reflection hump centered
around $\sim 30$ keV are perhaps the only significant observational
signatures of the presence of cold matter close to the event horizon
around accreting black holes in Active Galactic Nuclei (AGN) and
Galactic Black Hole Candidates (GBHC). This is why many theory papers
investigated X-ray reflection spectra from AGN and GBHCs in great
detail under the assumption that the matter is non-ionized or that the
density of the illuminated layer is constant (e.g., Lightman \& White
1988; George \& Fabian 1991, Ross \& Fabian 1993; Matt, Fabian \& Ross
1993, 1996; \zycki et al. 1994, and additional references in
NK). Basko, Sunyaev \& Titarchuk (1974); Kallman \& White (1989);
Raymond (1993); Ko \& Kallman (1994); \rozanska \& Czerny (1996)
relaxed the constant density assumption and all found that the thermal
ionization instability (Krolik et al. 1981) plays a central role in
establishing the equilibrium temperature and density profiles of the
X-ray illuminated gas. Nayakshin, Kazanas \& Kallman (2000; hereafter
NKK) extended results of these authors by providing accurate radiation
transfer for illuminating spectra appropriate for the inner part of
ADs in AGN and GBHCs. The results of NKK show that a self-consistent
gas density determination may provide valuable physical insights into
the problem that allow one to put tight constraints on AD theories.

As an example, Nayakshin \& Kallman (2000; NK hereafter) considered
the X-ray illumination problem in the three different AD geometries:
(1) the ``lamppost'' geometry, where the X-ray source is located above
the black hole at some height $h_x$; (2) full corona geometry (e.g.,
Liang \& Price 1979) and (3) the two-phase patchy corona model (e.g.,
Galeev, Rosner \& Vaiana 1979; Haardt, Maraschi \& Ghisellini 1994;
and Svensson 1996).  They pointed out that the reflected spectra and
correlations between the X-ray continuum and the atomic features, such
as the Fe K$\alpha$ line and the associated edge are very different
for these three geometries. Here we present the full disk spectra for
the lamppost and the flare models and broadly compare our theoretical
predictions to current observations of AGN. We find that a number of
observational facts rules out the lamppost model geometry and, at the
same time, supports the magnetic flare origin for the X-rays. (Note
that we do not discuss here the Advection Dominated Accretion Flows
(e.g., Ichimaru 1977; Rees et al. 1982; Narayan \& Yi 1994) or
modifications of this model due to winds (e.g., Blandford \& Begelman
1999; Quataert \& Gruzinov 2000) since these models are not expected
to work for many luminous AGN that have broad iron lines -- see the
recent review by Fabian et al. [2000]).

\section{Geometry of the X-ray source and the ionization state of the
disk}\label{sect:1rad}

Our main argument is based on two simple concepts. The first one has
to do with geometrical differences between the lamppost and the flare
models. In the former, the central X-ray source illuminates the disk
``evenly'' in the sense that the local illuminating flux at every
radius is the same as the azimuthal average of that flux. In the case
of flares, however, there may be many X-ray sources located anywhere
above the inner AD and only several disk height scales above it (e.g.,
Galeev et al. 1979; Haardt, Maraschi \& Ghisellini 1994; Nayakshin
1998). If the covering fraction of the disk by the flares is $f_c\ll
1$, then around the flare location (where most of the X-ray reflection
takes place), the illuminating flux if roughly $f_c^{-1}$ times larger
than the average of that flux. Therefore, for the same X-ray
luminosity, $L_x\sim L_d$, where $L_d$ is the disk bolometric
luminosity, the illuminating X-ray flux is $\fx\sim\fdisk$ and
$\fx\gg\fdisk$, for the lamppost and flare models, respectively, where
$\fdisk$ is the intrinsic disk thermal flux (see also Fig. 1 and
estimates in Nayakshin \& Dove 2000).

The second point is that the Compton temperature -- the maximum gas
temperature reached in the X-ray illuminated skin on the top of the
disk -- sensitively depends on the ratio $\fx/\fdisk$. Let $T_x$ be
the Compton temperature due to the illuminating X-radiation only,
which for typical Seyfert Galaxies spectra work out to be around
several to 10 keV. Then (e.g., Begelman, McKee \& Shields 1983),
\begin{equation}
T_c = {T_x J_x + \teff J_{\rm BB}\over
J_x +  J_{\rm BB}}\approx {T_x J_x\over
J_x +  J_{\rm BB}}\;,
\label{tcl}
\end{equation}
where we used $\teff\ll T_x$, and where $J_x$ and $J_{\rm BB}$ is the
angle integrated intensity of the X-rays and the soft disk flux,
respectively. Forgetting for the moment complications due to the angle
averaging of radiation and its transfer through the skin, $J_x/J_{\rm
BB} \sim \fx/\fdisk$, and hence the geometry of the X-ray source
directly influences $T_c$.

\begin{figure*}[t]
\centerline{\psfig{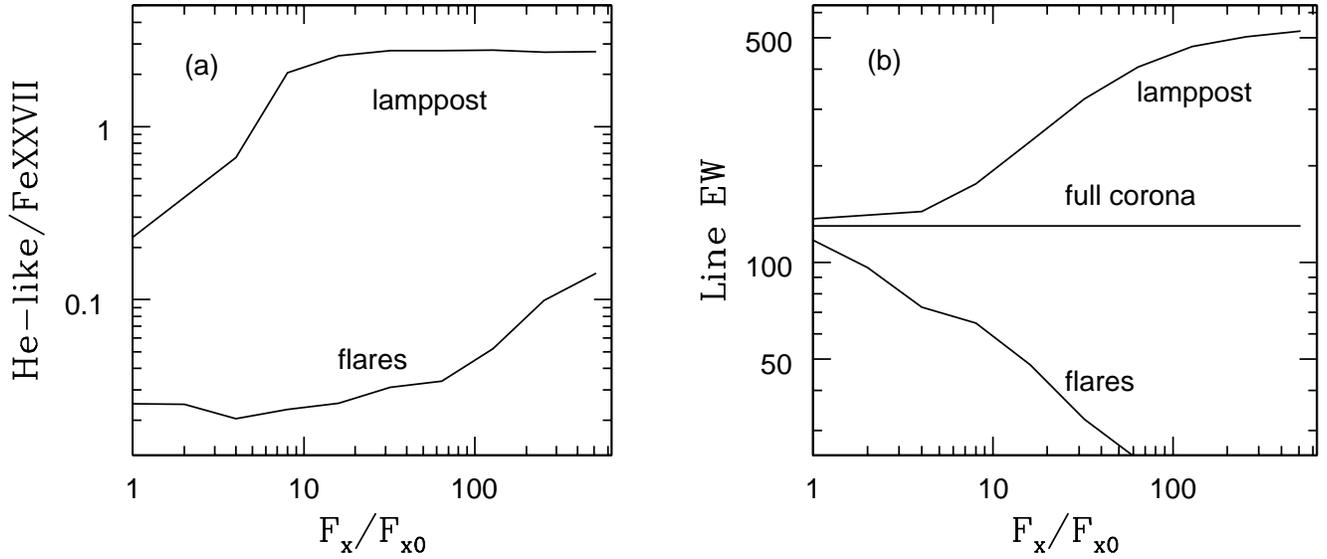}}
\caption{(a) -- Ratio of He-like iron to the completely ionized iron
in the skin for the lamppost and flare models as a function of the
X-ray ionizing flux for the runs presented in NK. (b) -- Behavior of
the Equivalent Width (EW) of iron K$\alpha$ line for the three
different models.}
\label{fig:ew}
\end{figure*}
NK made these arguments more precise by solving the X-ray illumination
problem at $R = 6 R_s$ ($R_s$ is the Schwartzchild radius) in the
lamppost geometry with the following ``reasonable'' parameters: $h_x =
6 R_s$, $\eta_x\equiv L_x/L_d = 1/5$, photon spectral index $\Gamma =
1.8$ and the exponential cutoff energy $E_c = 200$ keV. The gas
temperature on the top of the skin turns out to be only $k T \simlt
700$ eV. In contrast, in the magnetic flare case (NK assumed that
$\fx/\fdisk \sim 100$), the gas temperature on the top of the skin to
climb to about 4 keV (see Figs. 1 \& 5 in NK).  Not surprisingly, this
large difference in the gas temperature leads to as profound a
difference in the ionization structure of the gas.  For the lamppost
geometry, a large fraction of iron in the skin is in the form of
Helium-like iron.  For the flare case, on the other hand, the skin is
dominated by the completely ionized iron. Figure \ref{fig:ew}a shows
the ratio of the column depth of the He-like iron in the skin to that
of the completely ionized iron for the lamppost and the flare models
as a function of $F_x/F_{x0}$ for runs w1-w9 and h1-h7 presented in
NK, where $F_{x0}$ is the illuminating X-ray flux of the test with the
lowest value of $L_x$ (w1 and h1, respectively). For the lamppost,
He-like iron dominates by factor of few over completely ionized iron
in the skin, whereas for the magnetic flare case, He-like ions are
only few percent of the total iron line column depth in the skin.

The presence or absence of He-like iron in the skin makes it to play
two opposite roles for the reflected spectra. Because He-like iron has
much higher fluorescence yield than neutral-like iron below the skin,
the warm skin yields a higher iron line emissivity than the one from a
non-ionized matter. However, when the skin is completely ionized, it
obstructs penetration of the incident X-rays to the cooler layers
below where the Compton reflection hump and the fluorescent line are
formed, and further wash out these features by Compton scattering when
the line photons propagate through the skin to the observer (see
NKK). Figure \ref{fig:ew}b shows the dependence of the iron K$\alpha$
equivalent width (EW) on the X-ray flux. Clearly, the dependence of
the EW on $F_x$ is opposite for the two models. In both cases, the
Thomson depth of the skin increases with the X-ray luminosity (see
Nayakshin 2000), but because the skin in lamppost model is
He-like-dominated, this leads to an increase in the line's EW, whereas
for the completely ionized skin, it brings about a decrease in EW (the
cool matter below the skin is less and less ``visible'').

NK set several tests where they varied the ratio $\eta_x$ to determine
how high it should be to completely ionize the skin in the lamppost
case. It turned out that this ratio needs to be as high as $10$ or
larger, because otherwise the gas temperature on the {\em bottom} of
the skin falls far below the estimate given by equation 1 with
$J_x/J_{\rm bb} = \fx/\fdisk$ (see \S 6.3.1 and Figs. 13-15 in NK).
However, values of $\eta_x\simgt 1$ are in conflict with the fact that
many AGN have UV fluxes exceeding their X-ray fluxes (e.g., NGC~3516,
Edelson et al. 1999). Finally, NK showed that in the geometry of a
full corona sandwiching the cold disk, the additional weight/pressure
of the corona is so high that no ionized skin can form, and thus the
reflected spectra should resemble those of a neutral reflector. The EW
of the iron K$\alpha$ line is constant in this model (see
Fig. \ref{fig:ew}b).

\section{Full disk spectra}\label{sect:full}

In order to obtain full disk spectra, we calculate the local reflected
spectra, $I_k$, with the code of NKK and NK for 6 different radii $r_k
\equiv R/R_s = 3.5, 4.9, 7, 14, 35$ and 105. We assume that the
reflected spectrum for $3 < r < r_1$ is given by that for $r=r_1$, and
the one for $r > r_{6}$ is the one for $r=r_{6}$. Further, for radius
$r_k < r < r_{k+1}$, we define the reflected spectrum to be
\begin{equation}
I(r) = {\Delta r_{k+1} I_k + \Delta r_{k} I_{k+1}\over
r_{k+1} - r_k}\;,
\label{iofr}
\end{equation}
where $\Delta r_{k} = r - r_k$ and $\Delta r_{k+1} = r_{k+1} - r$. For
simplicity of notations, we dropped the dependency of $I_k$ on the
cosine of the viewing angle, $\mu$, and photon energy, $E$. In
practice, we calculate the reflected spectrum for 10 different viewing
angles $\theta_v$ that are measured from the normal to the disk, such
that $\mu_v \equiv \cos \theta_v =$ 0.05, 0.15, ..., 0.95. We then
interpolate in a linear fashion in $\mu_v$ (analogous to equation
\ref{iofr}) to obtain the reflected spectrum for an intermediate value
of $\mu_v$.

The full spectra are obtained via integrating over the disk surface.
To take into account the relativistic smearing in the Schwartzchild
geometry of a non-rotating black hole, we include gravitational
redshift in the photon energy and Doppler boosting due to Keplerian
disk rotation, but we neglect the ray bending in the vicinity of the
black hole. Since we only consider regions with $R > 3 R_s = 6
GM/c^2$, and since the maximum of the X-ray illuminating flux occurs
at yet larger radii, we believe that this approximation is adequate
for our study here.  Both models are calculated for a $3\times 10^8$
Solar masses black hole for dimensionless accretion rates: $\dm_j =
2^{-(2j+1)}$, where $j=0$, 1,..., 4 ($\dm = 1$ corresponds to the
Eddington luminosity).  We assume $\eta_x = 1/3$, $\Gamma = 1.8$ and
that the illuminating power-law extends to $E=200$ keV where it is
sharply cut.

\begin{figure*}[t]
\centerline{\psfig{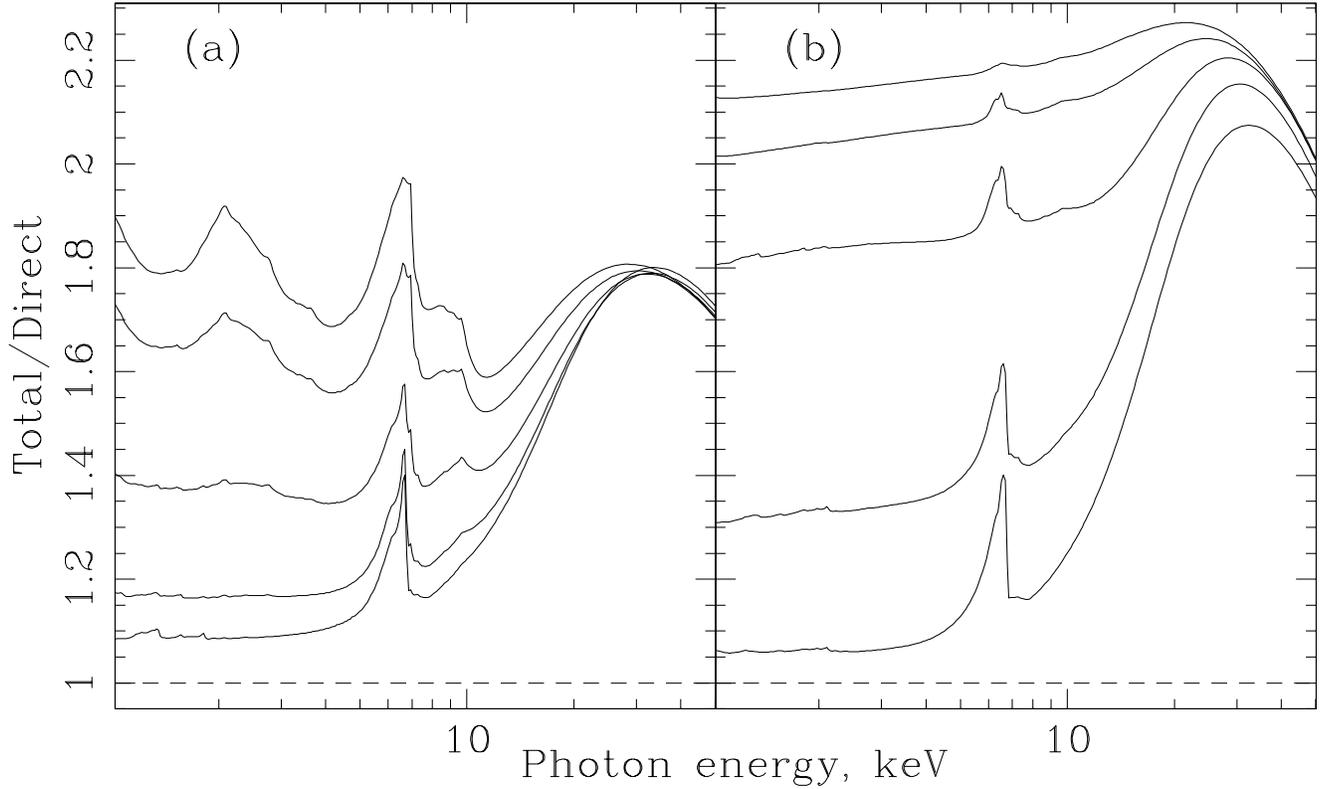}}
\caption{Ratio of the ``observed'' radiation intensity at an angle
$\theta = 30^{\circ}$ to that from the X-ray source(s) only for (a)
the lamppost model, and (b) the magnetic flare model. The
dimensionless accretion rates are $\dm_j = 2^{-(2j+1)}$, $j=0, ...4$,
from top to bottom. Note that the lamppost spectra are never
featureless because elements such as O, Mg, Si, S and Fe are not
completely ionized. On the other hand, the reflected spectra for the
magnetic flare model can be exact power-laws below $E\sim 20 keV$
because all these elements are completely ionized in the skin.}
\label{fig:spectra}
\end{figure*}

It is perhaps fair to say that moderate to high luminosity lamppost
spectra are rather unusual in that they display strong non-power-law
features at soft X-ray energies, very strong He-like iron line at
$\sim 6.7$ keV, and also a rather deep and high energy Fe absorption
edge (blended with Fe Ly continuum, see NK). These are generally {\em
not} observed in the spectra of real AGN or GBHCs. Note that the lower
normalization of the reflected spectrum of the lamppost case compared
with that for the magnetic flare case is due to two factors: (i) there
is a hole underneath the lamppost so that a part of the flux never
reflects; (ii) the X-ray continuum from magnetic flares is itself
beamed along the disk surface and smeared so that the observer
perpendicular to the disk observes less of the continuum flux than in
the lamppost case for the same {\em intrinsic} $L_x$.

Figure \ref{fig:spectra}b shows that below K$\alpha$ iron line
feature, the magnetic flare spectra are practically pure power-laws,
which is due to the fact that the skin is completely ionized.  As
shown by Done \& Nayakshin (2000), spectra for the magnetic flare
model {\em can} be fitted reasonably well by the standard single-zone
reflection amid a lower normalization for the reflection hump, i.e.,
these spectra do appear to be similar to the typical AGN and even hard
state GBHC spectra.

\newpage 
\section{Discussion}\label{sect:discussion}

A search for the iron line reverberation in response to continuum
variations was the goal of recent observations of MCG--6--30--15 (Lee
et al. 2000; Reynolds 2000) and NGC 5548 (Chiang et al. 2000). These
authors showed that whereas the continuum X-ray flux was strongly
variable on short time scales, the iron K$\alpha$ line flux stayed
roughly constant on these time scales and formally did not vary during
the whole observation, which is far longer than the light crossing
time of the innermost region. This seems to argue that the reflection
takes place very far from the black hole -- e.g., in the
putative molecular torus. However, the line profile is {\em broad},
indicating that most of its flux does come from the region rather
close to the black hole.
According to Figure \ref{fig:ew}b, the lamppost geometry and the full
corona case cannot explain the lack of iron line reverberation,
because that requires the EW of the line to drop with increase in the
X-ray flux approximately as EW$\propto \fx^{-1}$, whereas it increases
with $\fx$ for the lamppost model and stays constant for the full
corona geometry. On the other hand, the magnetic flare model can
potentially explain these interesting observations because the skin is
completely ionized in this model and thus the EW of the line decreases
with the X-ray flux.

Nandra et al. (1997) have shown that EW of the iron line monotonically
decreases with the increasing $L_x$ for a sample of AGN (the X-ray
``Baldwin'' effect). This study is corroborated by observations of
Vignali et al. (1999) who found that the available data for luminous
(high $z$) quasars indicate that the latter lack the reflection
component and also have no or weak iron lines.  This suggests that
either there is no cold matter near the X-ray source, or it is hidden
by a Thomson thick and completely ionized skin (as suggested by
Nayakshin 2000). The former possibility would be somewhat unexpected
because the currently popular AD theories either have the cold disk
going all the way to the last stable orbit black hole (e.g., magnetic
flares, lamppost) for all accretion rates, or suggest that the hot
part of the accretion flow diminishes with increase in $L$ and
eventually disappears (e.g., Esin, McClintok \& Narayan 1997; see also
\S 6.1 in Fabian et al. 2000). In other words, in all these theories
the luminous quasars are expected to have cold disk persisting down to
the last stable orbit, and hence the absence of the line and
reflection have to be explained by the ionization physics effects
(unless X-ray sources relativistically move away from the disk -- see
Beloborodov 1999 -- the more so the higher $L_x$ is).  Out of the two
models, the lamppost is clearly ruled out while the magnetic flare
model seems quite viable.

Further, the full disk spectra (Fig. \ref{fig:spectra}) show that the
lamppost model predicts many strong features due to Oxygen, Mg, Si and
other elements in soft X-rays, practically as prominent as those in
the iron recombination band, and an enormous EW for the iron line, and
a notably large iron absorption edge at $\sim 9$ keV.  Such spectra
are uncommon for real AGN, which again argues against the lamppost
model. On the other hand, crudely speaking, the reflected spectra of
the magnetic flare model is a combination of the mirror-like
reflection from the completely ionized skin and the cold-like
reflection hump and Fe K$\alpha$ line formed in the cold layers below
the skin. These spectra do appear to be neutral (for $\Gamma \simlt
2$, see NKK) and are reminiscent of spectra of real AGN and hard state
GBHCs (see Done \& Nayakshin 2000).  In addition, while some Seyfert 1
AGN do have extremely broadened iron lines like MCG--6--30--15
(e.g. NGC 3516; Nandra et al. 1999), others do not have the extreme
skewed lines expected from the very inner disk (e.g. IC4329a; Done,
Madejski \& \zycki\ 2000). As is clear from Figure 2b, the magnetic
flare model can explain both of these facts if we suggest that broad
iron line AGN are those with low $\dm \simlt 0.01$ or so, whereas
narrow iron line AGN accrete at larger $\dm$ such that the skin in
their inner disk completely destroys the broad line component. Note
that the lamppost model again fails here, because its skin increases
the line EW (see Fig. 2a).

Therefore, it emerges that the current observations indicate that the
ionized skin is indeed present on the top of accretion disks in AGN
and that it is {\em completely} ionized. The latter is only possible
if the illuminating X-ray flux is much larger than the disk flux,
which is most natural for accretion disks with magnetic flares
occurring above the disk.  The condition $\fx\gg \fdisk$ is in the fact
the basic assumption of the two-patchy phase model (same as the
magnetic flare model), because this requirement is central in
producing the X-ray continuum spectra as hard as those observed and
for a broad range of the parameter space (e.g., Haardt et al. 1994;
Stern et al. 1995; Poutanen \& Svensson 1996; Nayakshin \& Dove 2000).

{}

\end{document}